\documentclass[10pt,conference]{IEEEtran}
\usepackage{graphicx}
\usepackage{amsmath,amssymb}
\usepackage{booktabs}
\usepackage{cite}
\usepackage{url}
\usepackage{listings}
\usepackage{xcolor}
\usepackage[hidelinks]{hyperref}
\graphicspath{{images/}{./}}

\title{When Modernization Breaks Working Code:\\A Single-File Study of LLMs on Legacy PHP}

\newcommand{\authorcolumn}[3]{%
\begin{minipage}[t]{0.19\textwidth}
\centering
{\footnotesize\bfseries #1\par}
\vspace{3pt}
{\scriptsize\itshape #2\par}
{\scriptsize Bangladesh\par}
\vspace{2pt}
{\tiny #3\par}
\end{minipage}%
}

\author{
\makebox[\textwidth][c]{%
\authorcolumn
{Afridi Mahmud}
{Bangladesh University of\\Engineering and Technology}
{afridimahmud99@gmail.com}
\hfill
\authorcolumn
{Abid Hasan Khondaker}
{Bangladesh University of\\Engineering and Technology}
{abidhasan7116@gmail.com}
\hfill
\authorcolumn
{Sadif Ahmed}
{BRAC University}
{ahmedsadif67@gmail.com}
\hfill
\authorcolumn
{Md Nafiu Rahman}
{BRAC University}
{nafiu.rahman@gmail.com}
\hfill
\authorcolumn
{Rifat Shahriyar}
{Bangladesh University of\\Engineering and Technology}
{rifat@cse.buet.ac.bd}
}
}

\begin{document}
\maketitle

\begin{abstract}
Pasting a legacy file into an LLM and asking it to ``modernize this'' is an
easy shortcut. But is the result better than leaving the code alone? In this paper, we study
this single-file setting with five LLMs, each asked to modernize 100 WordPress
4.0 PHP files to PHP 8.3 while preserving behavior, with no project context or
execution feedback. We evaluate every output with 630 hidden behavioral checks
and 511 Rector-detected modernization findings. We compare against two
baselines: the unchanged originals and the rule-based tool Rector. Often, the
answer is no. Untouched, 83 files already pass every check; the best LLM leaves
only 59. Each model breaks 32--51 working files while repairing at most 10,
whereas Rector breaks 7 and repairs 6. Most of this damage is unnecessary. Only
37 findings (7\%) require a change under PHP 8.3, yet in compatible files with
no required change, the LLMs break 36--60\% (Rector: 8\%). The model with the
highest modernization score is no more compatible than its closest competitors.
Most regressed checks come from files that no longer load at all, often because
of type declarations PHP itself rejects. These are errors that need no project
context to avoid. Evaluations of LLM-based modernization should therefore
measure against a no-change baseline and report modernization and behavioral
preservation separately.
\end{abstract}
\begin{IEEEkeywords}
legacy modernization, large language models, PHP, behavioral preservation, empirical evaluation
\end{IEEEkeywords}

\section{Introduction}

Legacy code often keeps running after its language moves on, while looking
years out of date. Modernizing it means replacing outdated constructs
\emph{and} keeping the behavior callers depend on. LLMs make the first part
easy, but a modern-looking diff says nothing about the second.

\begin{figure}[t]
\centering
\includegraphics[width=.70\columnwidth]{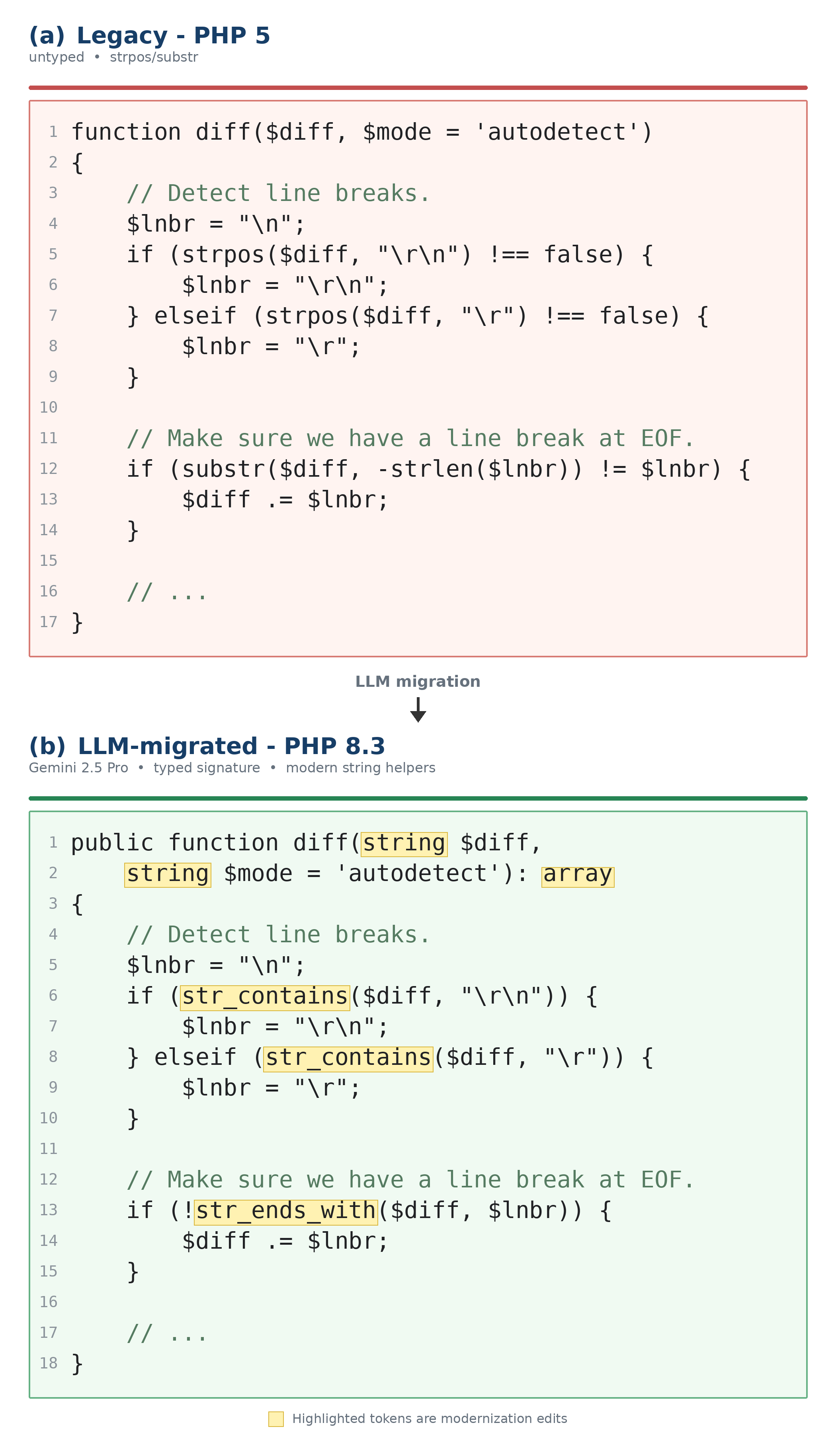}
\caption{Modernization of a legacy WordPress~4.0 diff method. (a) Original
method. (b) Gemini~2.5 Pro output generated from the target file alone.
Highlighted tokens show type declarations and modern string helpers.}
\label{fig:file_example}
\end{figure}

Figure~\ref{fig:file_example} shows why. Gemini~2.5 Pro adds type
declarations and modern string helpers to a legacy diff method, yet the
file fails one of its four hidden checks. For an unsupported diff mode, it
throws an \texttt{InvalidArgumentException} where the original returns the
\texttt{PEAR} error object that callers expect. Nothing in the source marks
this as a regression.

Migration benchmarks study whole repositories, often with build or test
feedback, or isolated functions and
snippets~\cite{liu2025migrationbench,fujii2026timemachine,wu2024versicode,amin2026jmigbench}.
They score the migrated code, but not against simply leaving it unchanged,
and they count required and optional changes alike. We study a common
intermediate setting: a developer pastes one complete file into a chat
assistant~\cite{xiao2024devgpt}, which sees the file's internal structure
but not its callers or dependencies. We evaluate the first answer, not an
agent's result after iterating on errors.

Five LLMs each modernize 100 WordPress~4.0 files to PHP~8.3 under a fixed
instruction requesting both modernization and preservation of inputs,
outputs, side effects, and call compatibility. We run 630 hidden checks and
compare against the unchanged originals and Rector~\cite{rectorDocumentation},
a rule-based modernization tool. We ask:
\begin{itemize}
\item \textbf{RQ1:} How do LLMs and Rector compare in preserving and
repairing tested behavior under single-file modernization?
\item \textbf{RQ2:} What do source-modernization scores reveal about
behavioral compatibility?
\item \textbf{RQ3:} What causes regressions after LLM modernization?
\end{itemize}

To our knowledge, this is the first modernization study to score LLM output
against the unchanged code's behavior while separating required from
optional changes. Under this design, every LLM ends with fewer compatible
files than the untouched originals (at most 59 versus 83), most of the
losses occurring in files that need no required change.

\section{Related Work}
LLM migration work spans transformation-by-example
(PyCraft)~\cite{dilhara2024pycraft} and industrial
workflows~\cite{ziftci2025migrating}. Benchmarks evaluate repository-level
migration (MigrationBench, TimeMachine-bench,
RepoMod-Bench)~\cite{liu2025migrationbench,fujii2026timemachine,li2026repomod},
version-sensitive snippets (VersiCode)~\cite{wu2024versicode}, and Java
functions (JMigBench)~\cite{amin2026jmigbench}. Pan et al.\ catalogue bugs
that LLMs introduce when translating between languages~\cite{pan2024lost},
and self-debugging uses execution feedback to repair generated
code~\cite{chen2024selfdebug}. Our study differs in three ways. We study
same-language modernization, where the unchanged original is itself a
runnable baseline. We evaluate the model's first answer for a whole file,
without project context or feedback. And we split source modernization
into required and optional changes, which aggregate scores conflate.

\section{Study Design}
Our design rests on one principle: every output is judged against what the
unchanged file already does. Originals, Rector output, and LLM output face
identical checks, and source modernization is measured separately from
behavior.

\subsection{Dataset and Baselines}
\begin{figure}[t]
\centering
\includegraphics[width=\columnwidth]{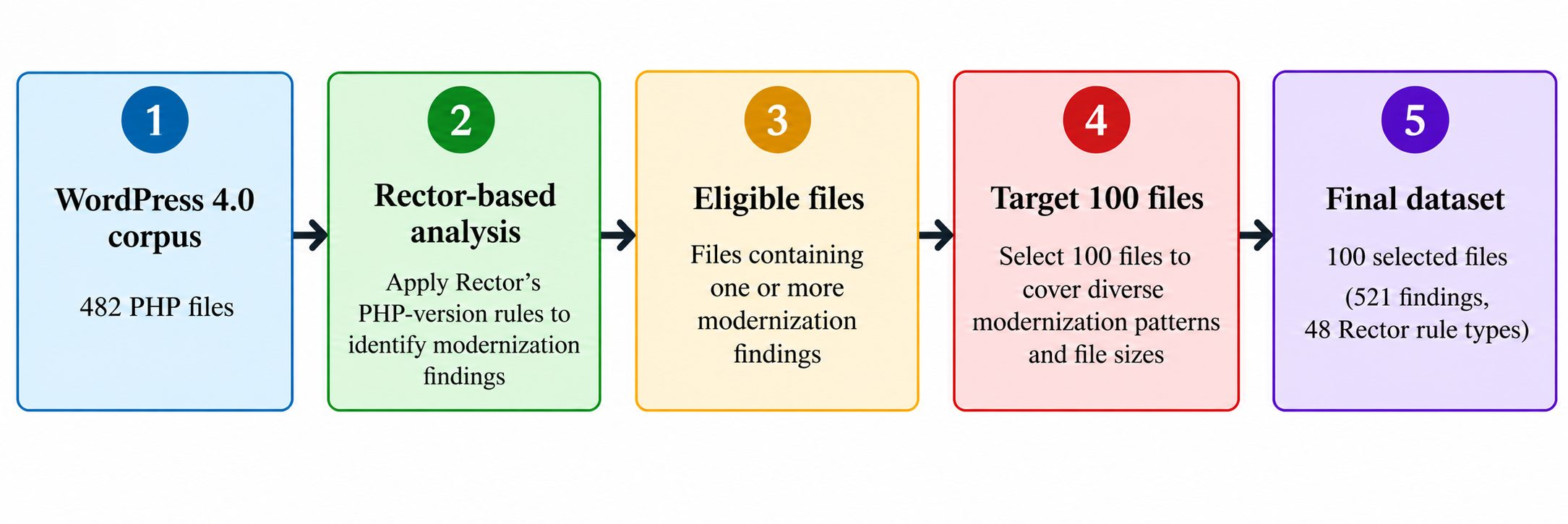}
\caption{Dataset selection from the 482-file WordPress~4.0 corpus. The 100 selected files contain 521 findings across 48 Rector rule types; excluding ten general code-quality findings leaves 511 findings across 44 PHP-version rule types for source-modernization scoring.}
\label{fig:dataset_selection}
\end{figure}

WordPress~4.0, released in 2014~\cite{wordpress40}, predates PHP~7 and
contains 482 PHP files. We analyze them with Rector~2.1.0, pinned through
Composer, using the \texttt{UP\_TO\_PHP\_83} level set; the 397 files with
findings are eligible. We deliberately favor pattern-rich files. A
deterministic procedure ranks eligible files within four size strata by
finding count, PHP-version families, and selected patterns, takes the
top-ranked files per stratum, and adds files until all 48 rule types are
covered (Figure~\ref{fig:dataset_selection}). The result spans 31 files of
at most 200 lines, 31 of 201--500, 26 of 501--1000, and 12 above 1000. This
choice gives every approach ample modernization opportunity. The sample is
purposive, so our inference concerns these 100 files, not WordPress or PHP
projects in general.

Rector plays two roles. As a detector, its PHP-version rules define what
needs modernizing: the selected files contain 521 findings across 48 rule
types, and excluding ten findings from four general code-quality or
dead-code rules leaves 511 findings across 44 PHP-version rule types for
scoring. The excluded rules remain in Rector's configured sets and can
therefore affect its output. As a baseline, Rector represents the
rule-based alternative a developer could run instead of an LLM. Like the
LLMs, it transforms one file at a time without execution feedback or manual
correction.

\subsection{Generation Protocol}
The instruction asks for modernization and preservation together, so any
regression violates an explicit requirement rather than an unstated one.
Its central task statement is:
\begin{quote}
\small Your task is to migrate this legacy PHP code up to PHP 8.3 standards
while preserving functional equivalence. Preserve observable behavior,
including inputs, outputs, side effects, and call compatibility.
\end{quote}
The prompt also assigns a senior PHP developer role, provides the complete
file, and requires code-only output between extraction markers. It was
fixed before any generated output was examined and is identical for all
five models (Table~\ref{tab:models}). Models receive no other source files,
tests, dependency information, errors, or runtime feedback.

\begin{table}[t]
\centering
\caption{Evaluated model API identifiers.}
\label{tab:models}
\footnotesize
\renewcommand{\arraystretch}{1.12}
\begin{tabular}{@{}ll@{}}
\toprule
Model & API identifier \\
\midrule
Claude Sonnet~4 & \url{claude-sonnet-4-20250514} \\
Gemini~2.5 Flash & \url{gemini-2.5-flash} \\
Gemini~2.5 Pro & \url{gemini-2.5-pro} \\
GPT-5 Codex & \url{gpt-5-codex} \\
Llama~3.3 70B Instruct & \url{meta-llama/llama-3.3-70b-instruct} \\
\bottomrule
\end{tabular}
\end{table}

GPT-5 Codex uses the OpenAI Responses API; Llama is served through
OpenRouter. Temperature is 0.2 except for GPT-5 Codex, whose API does not
expose that parameter; reasoning and thinking settings remain at provider
defaults. Requests allow up to 64,000 output tokens and 600 seconds. A
response is complete if it contains both markers and an entire file;
incomplete responses are regenerated, and the first complete response per
model and file is retained without editing. Regeneration counts were not
logged. A post hoc structural check shows that retained outputs keep
96--102\% of functions and 99--100\% of classes on average, with almost no
empty or stub bodies. Truncation or stubbing therefore cannot explain the
regressions we report.

\subsection{Behavioral Compatibility}
Hidden PHPUnit checks define the behavior to preserve. For each file, one
author selects public entry points, supplies inputs, and establishes
expected outputs and side effects from WordPress documentation and the
original implementation, before examining any generated file. A second
author, without access to model or Rector outputs, independently reviews
the behavior, expectations, and environment; disagreements are resolved by
discussion. Where an original fails a check, the oracle still encodes the
documented legacy behavior. For example, \texttt{IXR\_Client}'s constructor
uses \texttt{parse\_url} to assign host and port. The check expects
\texttt{example.com} and 8080; the original yields a null host under
PHP~8.3, whereas Rector's rewrite passes.

The 630 checks comprise 109 reachability checks, which test that a file
loads and runs without fatal errors or uncaught exceptions; 441 output or
state checks; and 80 fingerprint or artifact checks, which are sensitive to
implementation details. Section~\ref{sec:rq1} shows that our conclusions
hold without the fingerprint checks. A per-file environment supplies the
external WordPress functions, classes, global state, and data the selected
paths need, using WordPress~4.0 sources or hand-written stubs; we found no
reported regression attributable to harness or stub errors. Mean (median)
checks per file are 8.0 (7), 5.2 (5), 5.6 (5), and 6.4 (6) across the four
size strata. The checks cover selected paths, not complete behavior, but
any check that loads a language-invalid file detects its rejection
regardless of which path it targets.

Checks run under PHP~8.3 with \texttt{error\_reporting} set to
\texttt{E\_ALL}, and deprecation notices count as failures. Because this
policy applies equally to originals, it lowers the baseline and turns
every deprecation a model removes into a potential repair. If anything, it
favors modernization.

Each check is a preserved pass, regression, repair, or unrepaired failure
relative to the original. Preservation is the share of originally passing
checks that still pass; repair rate is the share of originally failing
checks that now pass. A file is \emph{compatible} if it passes all its
checks, \emph{broken} if it was compatible before modernization but not
after, and \emph{fixed} if the reverse. Because one fatal error fails every
check in a file, we report both file and check outcomes.

\subsection{Source Modernization and Failure Analysis}
Resolution measures what changed, not whether it should have. We reapply
the same Rector rules to each generated file. A finding is a (file, rule
type) pair and is resolved when that rule no longer fires on the generated
file, independently of behavior. Matching is by rule type, not location,
so deleting the affected code would also count as resolution; the
structural check above shows that such deletion is rare. Resolution is
resolved findings divided by 511. Findings in outputs Rector cannot parse
(3--12 files per model) count as unresolved, and newly introduced findings
are reported separately. We do not score Rector's resolution against its
own rules, which would be circular.

An aggregate score treats a change PHP~8.3 forces the same as a stylistic
one. We therefore classify findings as \emph{required}, whose constructs
fail or change behavior under PHP~8.3 (e.g., curly-brace string offsets,
PHP~4-style constructors); \emph{deprecated}, whose constructs run but
raise notices (e.g., null passed to non-nullable string arguments); and
\emph{optional}, whose constructs remain valid without notices (e.g., long
array syntax)~\cite{php80migration}. The rule-to-group mapping is in the
replication package. Every required finding is exercised by at least one
hidden check.

For RQ3, an automatic classifier assigns each regression to load-time
rejection, runtime type error, interface change, output mismatch, or other
failure, based on the first recorded PHP error or PHPUnit failure. Syntax
errors fail both measures: the file is rejected at load time, and its
findings count as unresolved.

\subsection{Statistical Analysis}
We compare file-level compatibility, which accounts for checks clustered
within a file, using two-sided exact McNemar tests~\cite{mcnemar1947} with
Holm--Bonferroni correction over all 21 pairwise comparisons among the five
LLMs, Rector, and the originals~\cite{holm1979}. As an effect size, we
report the paired difference in compatible files with a 95\% percentile
bootstrap interval (10,000 file resamples, seed~42). Both describe these
files under this protocol, not other projects.

\section{Results}
\subsection{RQ1: Behavioral Preservation and Repair}\label{sec:rq1}
Leaving the code alone is a strong baseline. The unchanged files pass 531
of 630 checks (84.3\%), and 83 of 100 files pass every check. A modernizer
therefore has far more working behavior to preserve (531 checks) than
failures to repair (99). Rector findings mark outdated constructs, not
necessarily broken behavior.

\begin{table*}[t]
\centering
\caption{Results on 100 files. The no-change baseline passes all checks for 83 files. Gap: compatible files lost relative to the originals (paired), with 95\% bootstrap interval. Broken and Fixed count files; Regressions counts checks. Preservation and repair use 531 originally passing and 99 originally failing checks; resolution uses 511 original findings.}
\label{tab:results}
\small
\setlength{\tabcolsep}{5pt}
\renewcommand{\arraystretch}{1.15}
\begin{tabular}{lrlrrrrrr}
\toprule
Approach & Compatible & Gap [95\% CI] & Broken & Fixed & Preservation & Repair & Regressions & Resolution \\
\midrule
Unchanged originals & 83 & -- & 0 & 0 & 100.0\% & 0.0\% & 0 & 0.0\% \\
\midrule
Claude Sonnet~4 & 59 & 24 [12, 35] & 32 & 8 & 76.3\% & 68.7\% & 126 & 53.2\% \\
Gemini~2.5 Flash & 50 & 33 [21, 45] & 40 & 7 & 62.7\% & 60.6\% & 198 & 53.4\% \\
Gemini~2.5 Pro & 55 & 28 [15, 40] & 38 & 10 & 72.3\% & 62.6\% & 147 & 72.8\% \\
GPT-5 Codex & 58 & 25 [13, 36] & 33 & 8 & 77.0\% & 58.6\% & 122 & 58.1\% \\
Llama~3.3 70B & 33 & 50 [40, 60] & 51 & 1 & 59.5\% & 21.2\% & 215 & 30.5\% \\
\midrule
Rector & \textbf{82} & 1 [$-6$, 8] & \textbf{7} & 6 & \textbf{97.7\%} & 62.6\% & \textbf{12} & -- \\
\bottomrule
\end{tabular}
\end{table*}

\paragraph{Every LLM loses more working behavior than it repairs}
No model matches the no-change baseline (Table~\ref{tab:results}). The
best, Claude Sonnet~4, ends with 59 compatible files against the
originals' 83. Across models, 32--51 compatible files are broken and only
1--10 incompatible files are fixed. Every LLM has significantly fewer
compatible files than the originals (Holm-adjusted $p\leq0.0021$), and no
bootstrap interval reaches zero. Among the LLMs, only Llama differs
significantly from the others: it trails Claude Sonnet~4, Gemini~2.5 Pro,
and GPT-5 Codex (adjusted $p\leq0.012$). No other pairwise LLM comparison
is significant after correction.

\begin{figure}[t]
\centering
\includegraphics[width=\columnwidth]{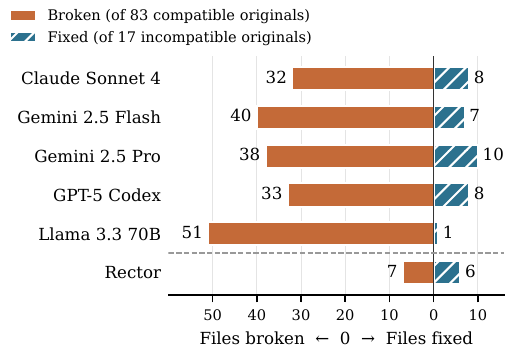}
\caption{File-level losses and gains relative to unchanged code. Broken files come from the 83 originally compatible files; fixed files come from the 17 originally incompatible files. Every LLM loses at least three times as many files as it repairs.}
\label{fig:transitions}
\end{figure}

\paragraph{Rector preserves behavior without standing still}
Rector's preservation does not come from inaction. It repairs 62
originally failing checks, as many as most LLMs (58--68 for four of the
five), while introducing only 12 regressions against their 122--215. Its
82 compatible files significantly exceed every LLM's result (adjusted
$p\leq0.0016$) and do not differ significantly from the originals' 83.
Four LLMs fix more whole files than Rector (7--10 versus 6), but each
breaks at least three times as many as it fixes
(Figure~\ref{fig:transitions}). Because Rector's resolution is not scored,
this is a claim about preservation, not modernization quality.

\paragraph{The gap survives both sensitivity analyses}
Excluding the 80 fingerprint checks leaves the originals and Rector at 83
and 82 compatible files and raises LLM counts by only 1--3
(Table~\ref{tab:robustness}), so the losses are not an artifact of
implementation-sensitive checks. A post hoc fallback that keeps a
generated file only if \texttt{php -l} passes, and otherwise keeps the
original, recovers 8--15 files per LLM. Even then, the best LLM reaches 69
compatible files, still 14 short of leaving the code alone. The remaining
14--38-file gap comes from output that passes lint but fails behavioral
checks.

\begin{table}[t]
\centering
\caption{Compatible files under two separate sensitivity analyses. No FP: exclude 80 fingerprint checks. Lint fallback: evaluate all 630 checks after substituting the original for lint-rejected output.}
\label{tab:robustness}
\footnotesize
\renewcommand{\arraystretch}{1.12}
\begin{tabular}{lrrr}
\toprule
Approach & All checks & No FP & Lint fallback \\
\midrule
Unchanged originals & 83 & 83 & 83 \\
Claude Sonnet~4 & 59 & 62 & 67 \\
Gemini~2.5 Flash & 50 & 53 & 65 \\
Gemini~2.5 Pro & 55 & 56 & 64 \\
GPT-5 Codex & 58 & 60 & 69 \\
Llama~3.3 70B & 33 & 36 & 45 \\
Rector & 82 & 82 & 82 \\
\bottomrule
\end{tabular}
\end{table}

\subsection{RQ2: Source Modernization versus Compatibility}\label{sec:rq2}
\paragraph{Higher resolution does not bring higher compatibility}
LLMs resolve 30.5--72.8\% of findings. Gemini~2.5 Pro resolves the most
(72.8\%), yet it ends with fewer compatible files (55) than Claude
Sonnet~4 (59) or GPT-5 Codex (58), which resolve 53.2\% and 58.1\%. These
differences are not significant, but the model that modernizes most shows
no preservation advantage.

\begin{table}[t]
\centering
\caption{Resolution by necessity under PHP~8.3. Introduced findings are counted separately.}
\label{tab:necessity}
\footnotesize
\setlength{\tabcolsep}{4pt}
\renewcommand{\arraystretch}{1.12}
\begin{tabular}{lrrrr}
\toprule
Model & Required & Deprecated & Optional & New \\
 & ($n=37$) & ($n=85$) & ($n=389$) & \\
\midrule
Claude Sonnet~4 & 78.4\% & 8.2\% & 60.7\% & 17 \\
Gemini~2.5 Flash & 78.4\% & 22.4\% & 57.8\% & 17 \\
Gemini~2.5 Pro & 78.4\% & 22.4\% & 83.3\% & 21 \\
GPT-5 Codex & 56.8\% & 32.9\% & 63.8\% & 20 \\
Llama~3.3 70B & 45.9\% & 9.4\% & 33.7\% & 18 \\
\bottomrule
\end{tabular}
\end{table}

\begin{figure}[t]
\centering
\includegraphics[width=\columnwidth]{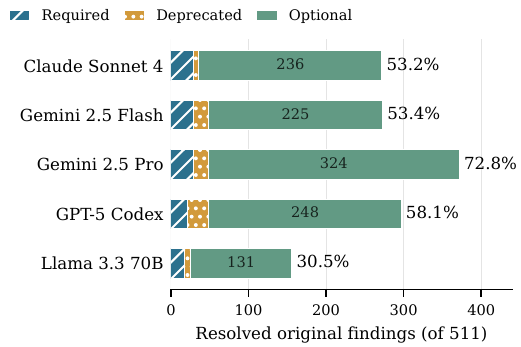}
\caption{Resolved findings by migration necessity. Labels inside optional segments give counts; bar-end labels give overall resolution out of 511 findings. Optional changes account for 82--87\% of resolved findings.}
\label{fig:composition}
\end{figure}

\paragraph{Most resolved findings are optional}
Only 37 findings are required (7\%), against 85 deprecated and 389
optional (Table~\ref{tab:necessity}). Optional findings account for
82--87\% of what each model resolves (Figure~\ref{fig:composition}).
Gemini~2.5 Pro's lead comes mainly from optional changes: it resolves
83.3\% of them, against 33.7--63.8\% for the other models, while matching
Claude and Flash on required findings (29 of 37 each). Every model
resolves deprecated findings least. Of these, 81 of 85 come from a single
rule guarding string arguments that might be null. Whether a value can be
null often depends on code elsewhere in the project, which the
single-file setting withholds.

\paragraph{Most broken files needed no required change}
Only 26 files contain a required finding, and 10 of them are compatible
unchanged. The other 74 contain only deprecated or optional findings, and
73 of them already pass every check. The two groups tell different
stories. Where change is required, the LLMs fix 6--9 files (Llama: 1) and
break 5--7, roughly breaking even; Rector fixes 6 and breaks 1. Where it
is not, the LLMs break 26--44 of the 73 compatible files (36--60\%) and
fix at most one, while Rector breaks 6 (8\%). For every LLM, most broken
files therefore needed no required change.

Nor does the amount of editing explain the losses. Among originally
compatible files, those in the upper half by changed-line share break in
50\% of cases against 43\% for the lower half (pooled), and the direction
varies by model. The kind of edit appears to matter more than its size,
and RQ3 traces many regressions to newly added declarations.

\subsection{RQ3: Causes of Regression}\label{sec:rq3}
Most regressions are not subtle. Between 66\% and 81\% of each LLM's
regressed checks occur in files with a failed reachability check: the
file fails to load or crashes on a tested path.

\begin{table}[t]
\centering
\caption{Check-level regression causes, with file-level counts below. C: Claude; F: Gemini Flash; P: Gemini Pro; G: GPT-5 Codex; L: Llama; R: Rector.}
\label{tab:causes}
\footnotesize
\setlength{\tabcolsep}{3.5pt}
\renewcommand{\arraystretch}{1.12}
\begin{tabular}{lrrrrrr}
\toprule
Cause & C & F & P & G & L & R \\
\midrule
Load-time rejection & 71 & 95 & 52 & 84 & 75 & 2 \\
Runtime type error & 18 & 20 & 20 & 12 & 23 & 2 \\
Interface change & 17 & 18 & 18 & 9 & 50 & 0 \\
Output mismatch & 19 & 43 & 33 & 16 & 48 & 8 \\
Other & 1 & 9 & 2 & 1 & 2 & 0 \\
Unclassified & 0 & 13 & 22 & 0 & 17 & 0 \\
\midrule
Total regressions & 126 & 198 & 147 & 122 & 215 & 12 \\
\midrule
Files with regressions & 36 & 45 & 41 & 36 & 56 & 9 \\
\quad with load-time rejection & 13 & 23 & 15 & 14 & 16 & 2 \\
\bottomrule
\end{tabular}
\end{table}

\paragraph{Load-time rejection accounts for the most regressed checks}
For every LLM, load-time rejection is the largest class of regressed
checks (Table~\ref{tab:causes}). Because a file that fails to load fails
all its checks, check-level counts amplify this class; at file level, it
affects 13--23 files per LLM, 29--51\% of their regressed files. Beyond
syntax errors, rejected declarations violate PHP's declaration rules or
break overrides. Claude Sonnet~4 and GPT-5 Codex each add a
non-\texttt{void} return type to \texttt{\_\_set()} in three files, and
both introduce a \texttt{?callable} property type, which PHP disallows.
In \texttt{class-IXR.php}, Claude adds a \texttt{bool} return type to
\texttt{IXR\_Client::query()} but not to its subclass override in the
same file. Each of these errors is detectable from the target file and
PHP's rules alone; none depends on the withheld project or the supporting
environment.

\begin{figure}[t]
\begin{lstlisting}
// Original (also retained by Rector)
public function __set($name, $value) {
    return $this->$name = $value;
}

// Claude Sonnet 4
public function __set(string $name,
                      mixed $value): mixed {
    return $this->$name = $value;
}
\end{lstlisting}
\caption{Observed method change in \texttt{class-wp-ajax-response.php}. The original and Rector output pass all nine checks. Claude's declared \texttt{mixed} return type violates PHP's requirement that a declared \texttt{\_\_set()} return type be \texttt{void}; its file fails to load, failing all nine checks.}
\label{fig:magic}
\end{figure}

Figure~\ref{fig:magic} shows the starkest case. Rector modernizes array
syntax elsewhere in the file but leaves the magic-method signature alone,
and all nine checks pass. Claude's \texttt{mixed} return type on
\texttt{\_\_set()} is invalid in any PHP~8 program, so the file fails to
load and all nine checks fail. Avoiding this error requires knowledge of
PHP, not of WordPress.

\paragraph{Other failures change accepted values, interfaces, or outputs}
Runtime type errors arise when new parameter or return types reject
values the original accepted. Interface changes include making legacy
public properties private and renaming or namespacing classes. Output
mismatches include returning integer \texttt{123} instead of string
\texttt{'123'}. Some captured subprocess output is truncated, hiding the
identifying text of fatal-error messages. This leaves 13 Flash, 22 Pro,
and 17 Llama regressions unclassified (Table~\ref{tab:causes}) but does
not imply that the environment caused them.

\paragraph{Rector shares one risk}
Rector is not immune. In two files it adds a \texttt{never} return type
that conflicts with overriding subclasses: on \texttt{WP\_Widget::widget()},
and on \texttt{Text\_Diff\_Op::reverse()}, whose subclasses sit in the same
file. Adding declarations without respecting the class hierarchy is
therefore a shared risk, but it appears in two Rector files against 7--11
files per LLM.

\section{Implications and Limitations}
\paragraph{Measure against the code you already have}
A modernized file is an improvement only if it beats the original.
Against 83 compatible originals, the best LLM's 59 is a net loss, yet its
53.2\% resolution would look like progress on its own. Migration
evaluations should report preserved behavior, regressions, and repairs
separately, and resolution scores should distinguish required,
deprecated, and optional changes.

\paragraph{Treat the first answer as a draft}
The lint fallback recovers 8--15 files per LLM at almost no cost but
leaves a 14--38-file gap, so lint is necessary, not sufficient. A staged
workflow (lint, load with dependencies, then check behavior) would catch
the load-time rejections that dominate regressed checks, including
invalid declarations that need no project context to detect. Execution
feedback could then drive repair~\cite{chen2024selfdebug}.

\paragraph{Scrutinize added declarations}
New type declarations are the most visible sign of modernization and a
recurring source of regressions, in LLM and Rector output alike. Tools
and reviewers should check added declarations against PHP's signature
rules and the class hierarchy before accepting them.

\paragraph{Construct validity}
Compatibility covers selected paths in an author-built environment, not
complete behavior. Three choices limit oracle bias: checks were written
before any generated output was examined, a second author reviewed them
without access to any output, and expected behavior, including for
originally failing checks, draws on WordPress documentation as well as
the legacy code. We found no reported regression attributable to harness
or stub errors, and the compatibility gap persists without fingerprint
checks. Every required finding is exercised, but untested paths may hide
further regressions as well as repairs. Counting deprecations as failures
applies equally to originals and, if anything, favors modernization.
Necessity classification involves author judgment, and first-error
classification may mask later defects.

\paragraph{Internal validity}
The instruction asks for ``PHP~8.3 standards,'' which invites optional
edits, but it also explicitly requires preserving call compatibility; the
regressions we report violate that explicit requirement. Whether a
narrower, required-changes-only instruction avoids them is untested.
Rector both selects files and detects findings, so the sample favors
patterns Rector recognizes. This may help Rector's repairs, but it cannot
explain LLM regressions, most of which occur in files needing no required
change. Regeneration of incomplete responses was not logged and may
affect longer files disproportionately, although the structural check
rules out truncation in retained outputs.

\paragraph{External validity and generation variability}
The results characterize five model versions under one instruction,
project version, and target runtime, with one generation per file. Newer
models may avoid specific errors, but the evaluation design applies
unchanged. Prompt sensitivity~\cite{sclar2024prompt} and generation
nondeterminism~\cite{ouyang2025nondeterminism} motivate repeated-generation
and alternative-instruction studies. A single repository and purposive
sample limit generalization beyond the selected files, and training-data
inclusion and OpenRouter's provider variation add uncertainty.
Repository-aware workflows extend the setting studied here.

\section{Conclusion}
Asked to modernize a legacy file while preserving its behavior, every
evaluated LLM left fewer working files than it started with. Unchanged,
83 of 100 WordPress~4.0 files pass every check; the best LLM leaves 59,
while Rector keeps 82. Modernization itself is not the problem: 16 of the
26 files with required findings fail unchanged, and the LLMs repair some
of them. The losses come mostly from changes PHP~8.3 did not require. The
LLMs break 36--60\% of compatible files that contain no required finding,
and most regressed checks come from files that no longer load, often
because of declarations PHP itself rejects. A modernization score alone
hides all of this. Evaluations should therefore measure against a
no-change baseline and report modernization, preservation, and repair
separately. A lint fallback recovers 8--15 files per LLM; whether
restricting LLMs to required changes closes the remaining gap is the
natural next question.

\begin{samepage}
\section*{Data Availability}
The anonymized replication package at \url{https://zenodo.org/records/20274676} contains the prompt, file list, raw and extracted outputs, test harness and stubs, rule-necessity classification, and the statistical and reanalysis scripts with their CSV outputs.
\par
\end{samepage}

\bibliographystyle{IEEEtran}
\bibliography{references}
\end{document}